\documentclass[useAMS,usenatbib]{mn2e}
\usepackage{graphicx}
\usepackage{mathptmx}
\usepackage{dcolumn}
\newcolumntype{d}{D{.}{.}{-1}}

\def\SNR(#1.#2)#3(#4.#5){{G#1${\cdot}$#2$#3$#4${\cdot}$#5}}
\graphicspath{{.}}
\newcounter{todo}
\renewcommand\thetodo{\Alph{todo}}
\def\todo#1{\addtocounter{todo}{1}[[\thetodo: #1]]\strut\vadjust{%
\kern-\dp\strutbox{\vtop to \dp\strutbox{\baselineskip\dp\strutbox\vss\rlap{%
\hskip\hsize\ \rm{$\leftarrow$\thetodo}}\null}}}}
\def\note#1{\strut\vadjust{\kern-\dp\strutbox{\vtop to \dp\strutbox{%
\baselineskip\dp\strutbox\vss\rlap{\hskip\hsize\ {\tiny\rm #1}}\null}}}}
\title[\SNR(64.5)+(0.9), a new shell supernova remnant with unusual central emission]
{\SNR(64.5)+(0.9), a new shell supernova remnant with unusual central emission\thanks{We request that any reference to this paper cites ``AMI Consortium: Hurley-Walker et al. 2009"}}

\label{firstpage}

\author[Hurley-Walker et~al.]
{AMI CONSORTIUM:
Natasha~Hurley-Walker$\thanks{Issuing author; E-mail: nh313@mrao.cam.ac.uk}$,
 Matthew~L.~Davies, \newauthor
Thomas M. O. Franzen, Keith~Grainge, D.~A.~Green, Michael~P.~Hobson, \newauthor
Anthony~Lasenby, Guy~Pooley, Carmen~Rodr\'{\i}guez-Gonz\'alvez, \newauthor
Richard~D.~E.~Saunders, A.~M.~M.~Scaife, Paul~F.~Scott, Timothy Shimwell, \newauthor 
David~Titterington, Elizabeth~Waldram and Jonathan~T.~L.~Zwart\\
Astrophysics Group, Cavendish Laboratory, 19 J. J. Thomson Avenue,
Cambridge CB3 0HE\\
}

\date{Accepted ---; received ---; in original form \today}
\pagerange{\pageref{firstpage}--\pageref{lastpage}}
\pubyear{2009}

\begin{document}
\maketitle

\begin{abstract}
We present observations between 1.4 and 18\,GHz confirming that \SNR(64.5)+(0.9) is new Galactic shell
supernova remnant, using the Very Large Array and the 
Arcminute Microkelvin Imager. The remnant is a shell $\simeq8$\arcmin\, in diameter
with a spectral index of $\alpha=0.47\pm0.03$ (with $\alpha$ defined such that flux density $S$
varies with frequency $\nu$ as $S\propto\nu^{-\alpha}$).
There is also emission near the centre of the shell, $\simeq1$\arcmin\ in extent, with
a spectral index of $\alpha=0.81\pm0.02$.
We do not find any evidence for spectral breaks for either source within our frequency range.
The nature of the central object is unclear and requires further investigation, but we
argue that is most unlikely to be extragalactic. It is difficult to avoid the conclusion
that it is associated with the shell, although its spectrum is very unlike that of known pulsar
wind nebulae.
\end{abstract}

\begin{keywords}
  supernova remnants -- radio continuum: ISM -- radiation
  mechanisms: non-thermal
\end{keywords}

\section{Introduction}

\begin{figure}
\centerline{\includegraphics[width=6.5cm,height=6.5cm,keepaspectratio,angle=-90]{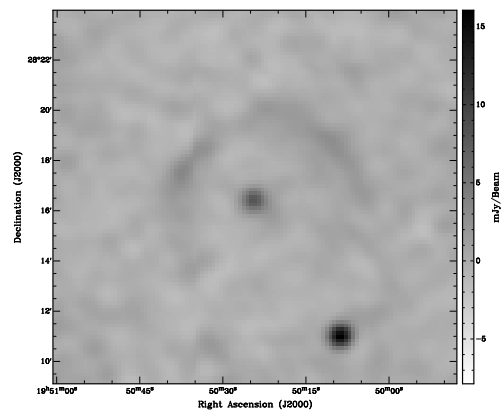}}
\caption{NVSS image of \SNR(64.5)+(0.9); the resolution is 45 arcseconds and the map noise is 3~mJy~beam$^{-1}$.
\label{fig:nvss}}

\end{figure}

While examining an NRAO VLA Sky Survey (NVSS) image \citep{1998AJ....115.1693C} of the region surrounding the Galactic supernova remnant
\SNR(63.7)+(1.1), an arc of extended emission
was noticed near RA$ = 19^{h} 50^{m} 24^{s}$, Dec$ = 28^{\circ} 16\arcmin 25''$
(see Fig. \ref{fig:nvss}).
As it resembles part of a ring 
and lies in the Galactic plane, it was identified as a possible supernova remnant (SNR). \citealt{2006A&A...455.1053T}
also noticed this structure in Canadian Galactic Plane Survey 
(CGPS; \citealt{2003AJ....125.3145T}) data with similar sensitivity to but poorer resolution than the NVSS data,
and proposed it as a possible SNR. It has now been 
observed with the VLA and the newly commissioned Arcminute Microkelvin Imager (AMI; see \citealt{2008MNRAS.391.1545Z}).
These observations are described in Section 2, and calibration and data reduction in Section 3.
Results are presented in Section \ref{section:results}, and we analyse and discuss all available data
in Section \ref{section:discussion}, confirming \SNR(64.5)+(0.9) as a new Galactic SNR.

\section{Observations}

\begin{table*}
\centering
\caption{Observations of \SNR(64.5)+(0.9).\vspace{0.2cm}
\label{tab:obslist}}
\begin{tabular}{l|ccccc}
\hline
Telescope & AMI SA & AMI LA & \multicolumn{3}{c}{VLA}\tabularnewline
\hline
Date & 2007 October & 2008 October & \multicolumn{3}{c}{2008 July}\tabularnewline
Flux Calibrator & 3C286 & 3C48 & \multicolumn{3}{c}{3C48} \tabularnewline
Phase Calibrator & J2023+3153 & J2023+3153 & \multicolumn{3}{c}{J1925+211}\tabularnewline\cline{4-6}
Frequency/ GHz & 14--18 & 14--18 & 1.43 & 4.86 & 8.46 \tabularnewline
Observation(s) Length & 16 hours & 10 hours & 40 minutes & 40 minutes & 20 minutes\tabularnewline
Synthesised beam FWHM & 3--2\arcmin & 1--0.3\arcmin & 0.75\arcmin & 0.22\arcmin & 0.12\arcmin \tabularnewline
Primary beam FWHM & 22--16\arcmin & 5.6--5.0\arcmin & 32\arcmin & 9.3\arcmin & 5.3\arcmin \tabularnewline
\hline
\end{tabular}
\end{table*}

The position of the source close on the sky to the ring centre
was chosen as the pointing centre for observations
with the VLA and AMI.
AMI is a dual set of interferometric arrays located at the Mullard Radio
Astronomy Observatory, Lord's Bridge, Cambridge, UK. The AMI Small Array (SA) consists
of ten 3.7-m diameter equatorially-mounted dishes with a baseline range of
$\simeq 5$ -- 20\,m, while the AMI Large Array (LA) has eight 12.8-m-diameter dishes
with a baseline range of $\simeq 20$ -- 100\,m. Both arrays observe I+Q in the band 12--18\,GHz, each with
system temperatures of about 25~K.

The backends are analogue Fourier
transform spectrometers, from which the complex signals in each of eight
channels of 750-MHz bandwidth are synthesised, and the signals in the synthesised channels
are correlated at the $\simeq 10$ per cent level. In practice, the
lowest two frequency channels are generally not used due to a poor correlator response in
this frequency range and, currently, interference.

Frequencies, resolutions, dates, calibrators and observing times for
each observation are shown in Table~\ref{tab:obslist}. The AMI SA observed a single pointing
while the AMI LA used a seven-point raster mode with 4\arcmin\,spacing. 
J2023+3153 (RA$ = 20^{h} 23^{m} 19^{s}.0$, Dec$ = 31^{\circ} 53\arcmin 02''.3$)
was chosen as a phase calibrator from the Jodrell Bank VLA Survey
(JVAS; \citealt{1992MNRAS.254..655P, 1998MNRAS.293..257B,
1998MNRAS.300..790W}) on the basis of its proximity and flux density (1.1\,Jy at 15\,GHz). 

During the time of our VLA observations, about half of the antennas of the VLA were upgraded Extended VLA (EVLA) antennas.
This necessitated an extra step of calibration on VLA--EVLA baselines.
The telescope was in D-array, and we used the L, C and X band receivers at centre frequencies of 1.43, 4.86 and 8.46 GHz respectively. J1925+211 (RA$ = 19^{h} 25^{m} 59^{s}.6$, Dec$ = 21^{\circ} 06\arcmin 26''.2$) was
selected from the VLA Calibrator Manual\footnote{http://www.aoc.nrao.edu/$\sim$gtaylor/csource.html}
as the phase calibrator at 1.43, 4.86 and 8.46 GHz in D-array;
it has an assumed flux density of 1.3, 1.5 and 1.0\,Jy in each 
of these bands, respectively.

\section{Calibration and Data Reduction}

The AMI data reduction was performed using our software tool \textsc{reduce}. This is used to apply path-compensator and path-delay corrections, to flag interference,
shadowing and hardware errors, to apply phase and amplitude calibrations and to
Fourier transform the correlator data readout to synthesise the frequency channels,
before outputting to disk in $uv$-\textsc{FITS} format suitable for imaging in \textsc{aips}.

Flux calibration was performed using short observations of 3C48 and 3C286 near
the beginning and end of each run, with assumed I~+~Q flux densities for these sources in
the AMI channels consistent with \citet{1977A+A....61...99B} (see Table~\ref{tab:Fluxes-of-3C286}). 
As \citeauthor{1977A+A....61...99B} measure I and AMI measures I~+~Q, these flux densities
include corrections for the polarization of the sources derived
by interpolating from VLA 5-, 8- and 22-GHz observations.
After phase calibration, the phase of AMI over one hour is generally
stable to $5^{\circ}$ for channels 4--7, and to $10^{\circ}$ for channels 3 and 8.

\begin{table}
\centering
\caption{Assumed I~+~Q flux densities of 3C286 and 3C48 over the commonly-used AMI band.  \label{tab:Fluxes-of-3C286}}
\begin{tabular}{ccccc}\hline
 Channel & $\nu$/GHz & $S^{{\rm {3C286}}}$/Jy & $S^{{\rm {3C48}}}$/Jy \\ \hline
 3 & 14.2 & 3.61 & 1.73 \\
 4 & 15.0 & 3.49 & 1.65 \\
 5 & 15.7 & 3.37 & 1.57 \\
 6 & 16.4 & 3.26 & 1.49 \\
 7 & 17.1 & 3.16 & 1.43 \\
 8 & 17.9 & 3.06 & 1.37 \\ \hline
\end{tabular}
\end{table}

The system temperatures of each AMI antenna are
continously monitored using a modulated noise signal injected at each
antenna; this is used to continuously correct the amplitude
scale in a frequency-independent way. The overall consistency of the
flux-density scale is estimated to be better than 5 per cent.

Since the AMI antennas are sensitive to I+Q and are equatorially mounted,
this polarization is fixed on the sky during the observation.
Q may be positive or negative, but is expected to be small
when integrated over the whole source.

VLA data reduction was performed entirely within \textsc{aips}. The VLA--EVLA 
baselines were also calibrated using \textsc{blcal} as described in the
guidelines for post-processing EVLA data in \textsc{aips}\footnote{http://www.vla.nrao.edu/astro/guides/evlareturn/postproc.shtml}. This corrects for closure errors on these baselines caused by non-matched EVLA and VLA bandpass shapes.

\begin{figure}
\centerline{\includegraphics[width=7.5cm,height=7.5cm,keepaspectratio,angle=-90]{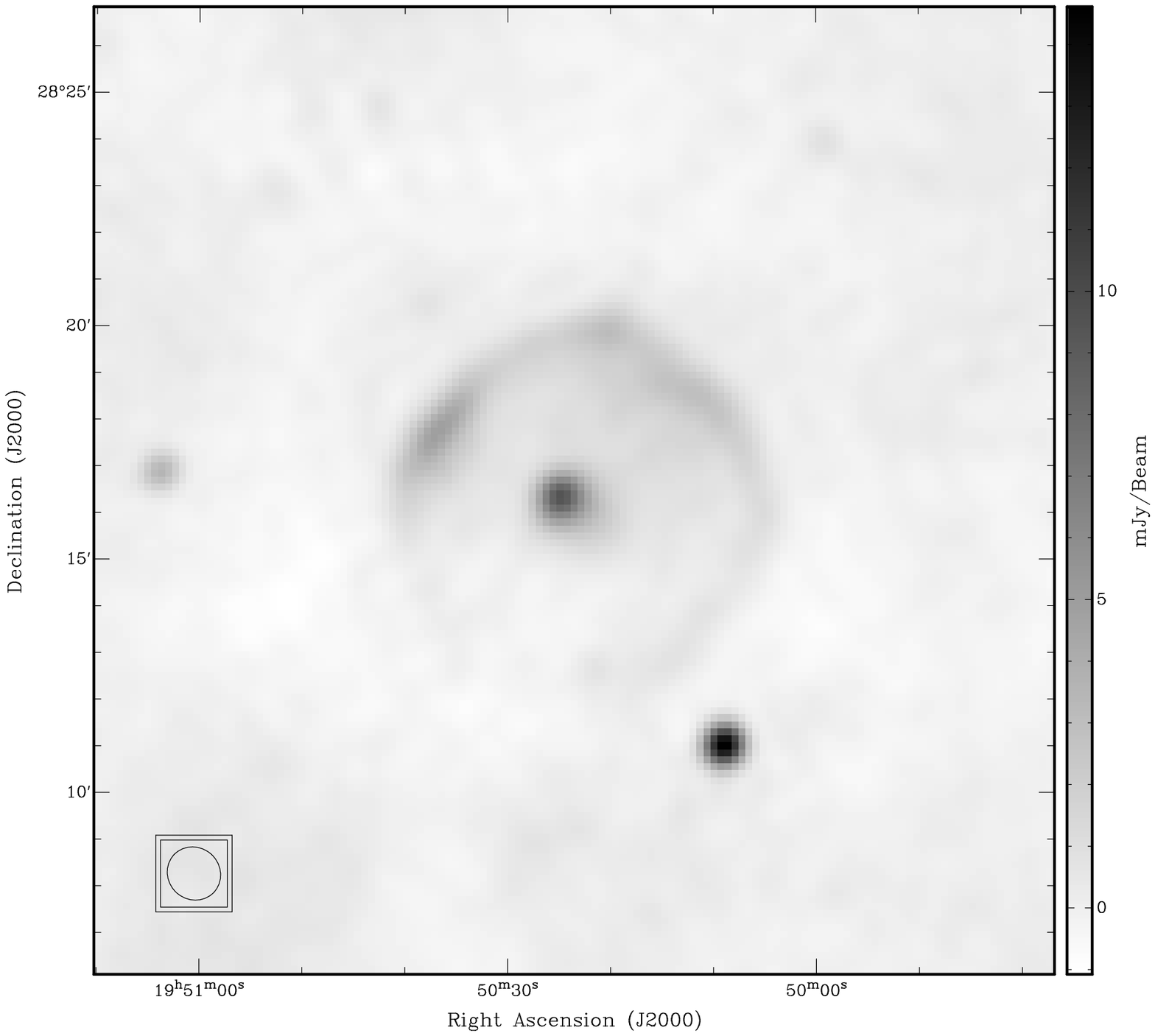}}
\centerline{\includegraphics[width=7.8cm,height=7.8cm,keepaspectratio,angle=-90]{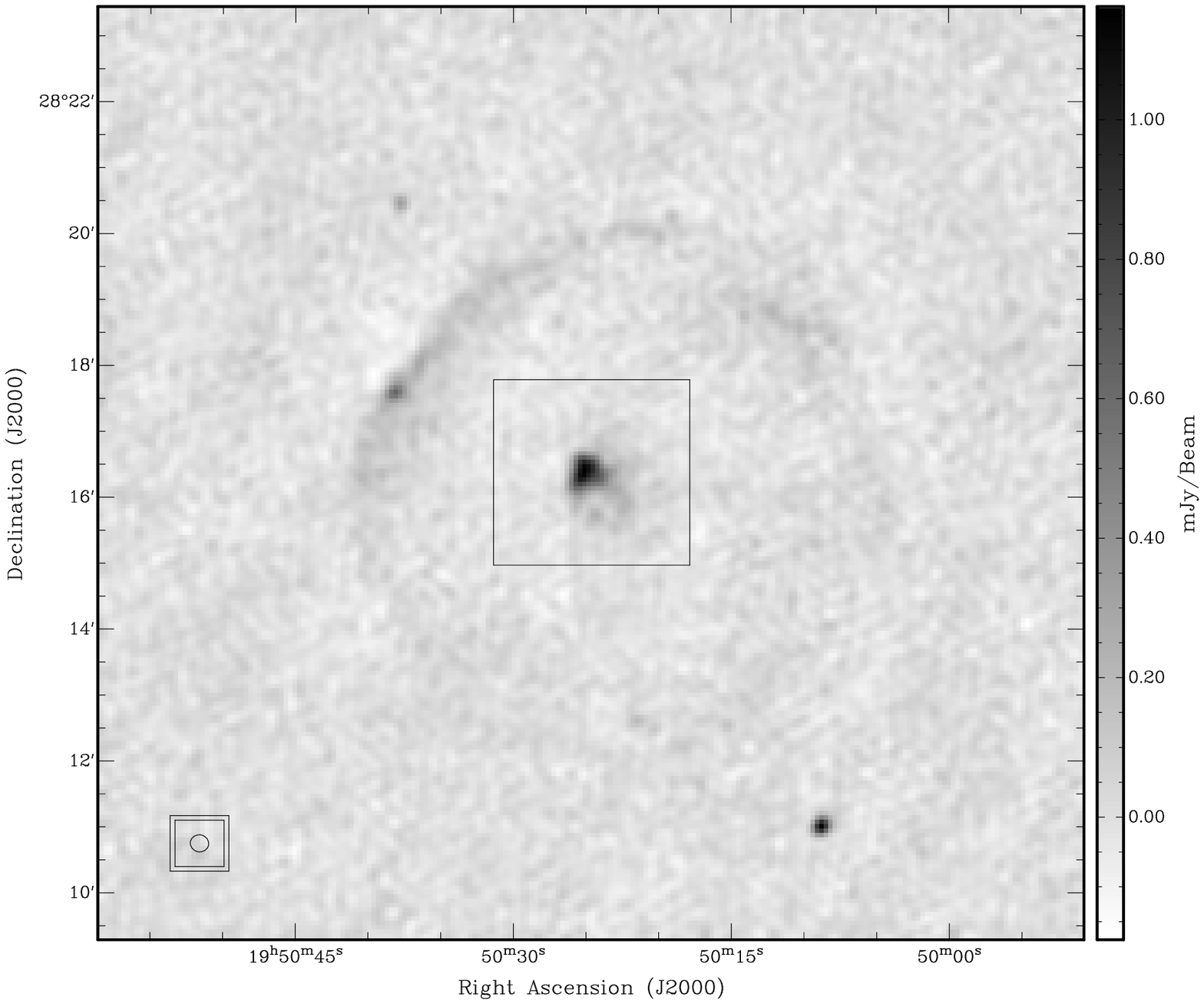}}
\centerline{\includegraphics[width=8.1cm,height=8.1cm,keepaspectratio,angle=-90]{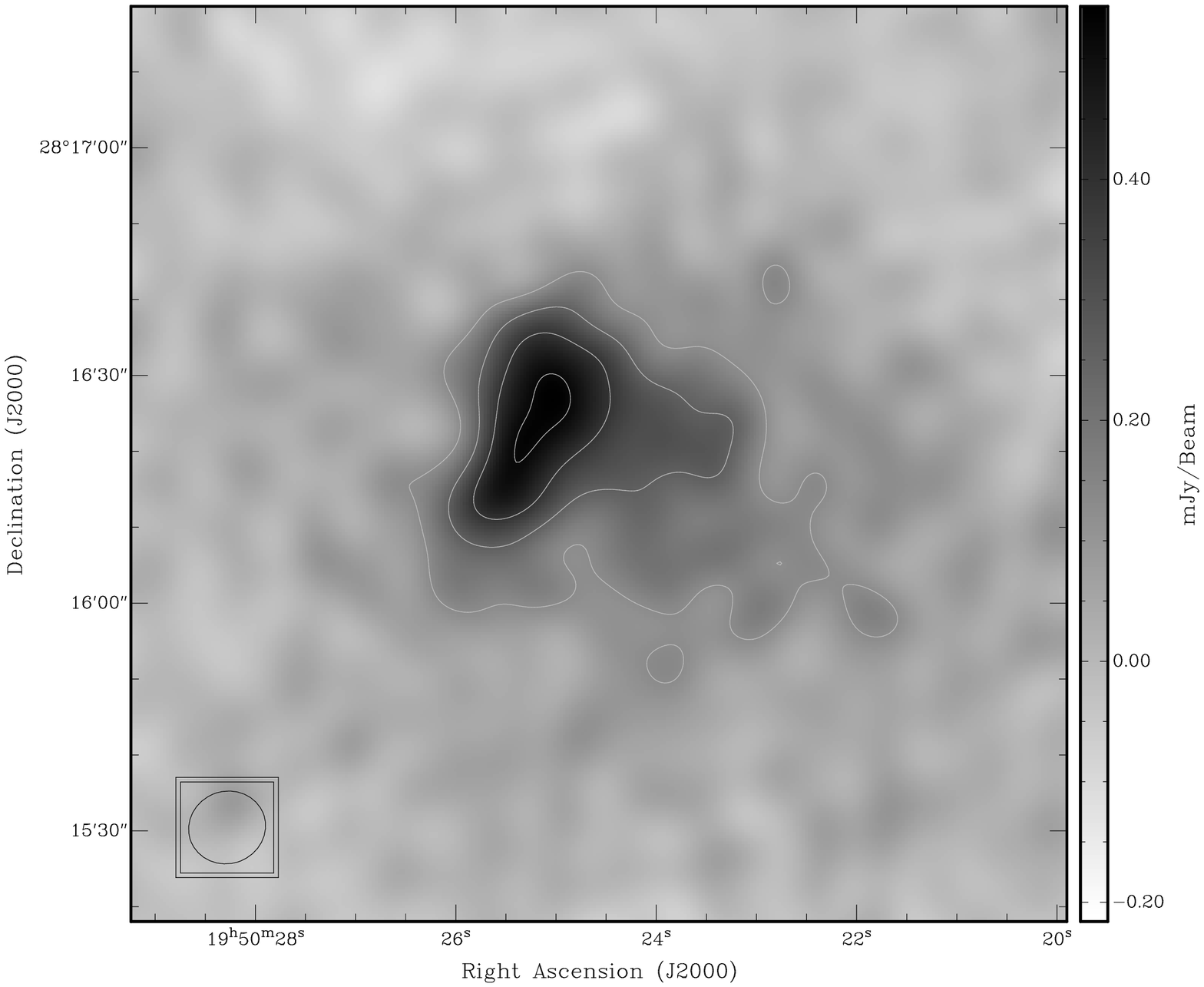}}
\caption{VLA images of \SNR(64.5)+(0.9). Top: 1.43\,GHz, $\sigma_{rms}=110~\mu$Jy~beam$^{-1}$, uniform weighting;
middle: 4.86\,GHz, $\sigma_{rms}=38.5~\mu$Jy~beam$^{-1}$, uniform weighting;
bottom: 8.46\,GHz, $\sigma_{rms}=43.5~\mu$Jy~beam$^{-1}$, natural weighting with three-sigma contours. The
box in the middle panel indicates the area that is shown in the lower panel.
In this and subsequent images, the FWHM of the \textsc{clean} restoring beam of the observation is shown as an ellipse
inside a box on the lower left.\label{fig:VLA}}
\end{figure}

Maps were made using \textsc{imagr} in \textsc{aips} from each channel of the AMI SA, AMI LA
 and the VLA (Fig.~\ref{fig:VLA}). Combined-channel maps of the
AMI SA and AMI LA observations are shown in Fig. \ref{fig:AMISA} and Fig. \ref{fig:LA-raster},
respectively.
The VLA and AMI SA images shown are not corrected for the primary beams of the
telescopes, but when measuring flux densities this correction is made.
In order to produce a raster map, the AMI LA images are of course primary-beam corrected.
The signal-to-noise of the 1.43-GHz image was large enough
that self-calibration from the first non-negative \textsc{clean} components was used to enhance the image.

\section{Results}\label{section:results}

We note that the maps show two sources of emission: a ring-like
structure surrounding a slightly extended, apparently central object.
In those maps that do not resolve the ring completely, the emission
from these two sources may be confused.

As we are using interferometers sensitive to a limited range of angular scales,
it is not possible to obtain accurate flux densities for structures on all scales from all of the observations.
Only the VLA 1.43-GHz and AMI SA observations have the required short baselines to provide sensitivity
to the extended ring emission. The AMI SA does not have the required resolution to measure
the flux density of the central emission without confusion with the ring.

To measure flux density, we adopt the fitting method of \citet{2007BASI...35...77G}, using
the program \textsc{fitflux}.
In this method, a flux density is fitted by drawing a polygon
around the object and fitting a tilted plane to the pixels around the
edges of the polygon. The tilted plane is then removed from the image
before integrating the emission within the polygon. This method is sensitive to the
fitting area selected but only at the one per cent level, which is included in the estimate
of the error. This method allows us to remove contribution from the background,
especially the contribution from the ring in the case of measuring the flux density of the
central emission.

The main sources of error on the flux density measurements are the thermal noise $\sigma_{\rm{rms}}$
and the error on the flux calibration
$\sigma_{\rm{S}}$. Long-term measurements of the AMI flux calibration show it to have an r.m.s. error
of three per cent. The VLA flux calibration is performed using \textsc{getjy} which scales the fluxes of the field
and phase calibrator using the flux density of 3C48 consistent with \citeauthor{1977A+A....61...99B}. This process has an error of less than
one per cent for each of the frequency channels. Therefore to obtain the error on a flux density measurement,
the thermal noise of each map is added in quadrature with the flux calibration error of the telescope used, and
a one per cent error when \textsc{fitflux} is used.

We now discuss the results from each observation in turn.

\subsection{VLA 1.43\,GHz}

Fig.~\ref{fig:VLA} shows that this object
is clearly a shell SNR, brighter in the north-east. Interestingly, the central emission
is also brightest in this direction. The full ring 
structure is clearly discernable in this map, as compared to
the partial structure shown in the NVSS map (Fig. ~\ref{fig:nvss}). This is due
to the enhanced $uv$ coverage and lower noise of the longer pointed observation.

\textsc{fitflux} was used to obtain the flux density of the central object as $17.1\pm0.8$\,mJy,
and the flux density of the entire remnant as $119\pm5$\,mJy.
The former was subtracted from the latter to
obtain the flux density of the ring as $102\pm5$\,mJy.

\subsection{VLA 4.85\,GHz}

At 4.85\,GHz, the ring is resolved out to the extent that it is not possible
to extract a reliable estimate of its integrated flux density. However, the 
structure of the emission close to the ring centre becomes more clear;
its integrated flux density is listed in Table~\ref{tab:centreflux}.

The north-east section of the ring shows possible contamination with a point source. Using the \textsc{aips}
routine \textsc{slice}, a cross-sectional profile across this source, perpendicular to a tangent along the ring at this point, was
produced in order to measure the flux levels. The base flux level from the ring is $220~\mu$Jy~beam$^{-1}$
and the peak point source flux is $620~\mu$Jy~beam$^{-1}$. The width of the source profile at $420~\mu$Jy~beam$^{-1}$
is 12\arcsec, which is the resolution of the telescope at this frequency. Therefore the source
is unresolved and we postulate it is a background contaminating source. \textsc{fitflux} gives
a flux density estimate of $0.41\pm0.04$\,mJy for this source.

\subsection{VLA 8.46\,GHz}

At 8.46\,GHz, the ring is almost entirely resolved out. However we obtain good information
on the structure of the near-central emission, as shown in the lowest panel of Fig. \ref{fig:VLA}.
\textsc{fitflux} gives a measurement of the flux density of this source, shown in Table~\ref{tab:centreflux}.

\subsection{AMI 14--18\,GHz}

\begin{figure}
\centerline{\includegraphics[width=7.4cm,height=7.4cm,keepaspectratio,angle=-90]{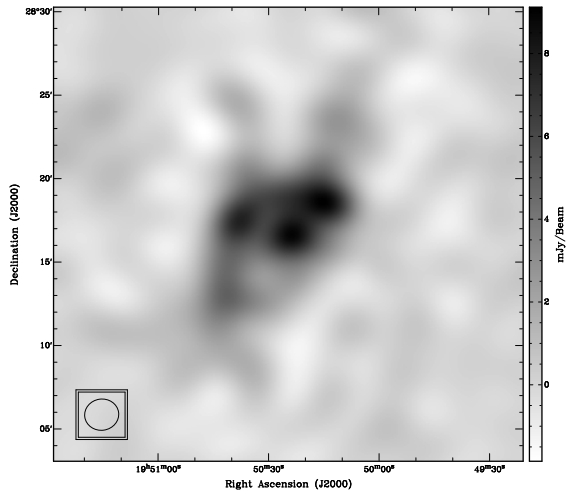}}
\caption{AMI SA combined-channel (14--18\,GHz) image of \SNR(64.5)+(0.9);
the map noise is $192~\mu$Jy~beam$^{-1}$.\label{fig:AMISA}}
\end{figure}

The AMI SA, due to its sensitivity to larger angular scales, picks up much more
large-scale Galactic emission in the region (see Section~\ref{section:lit_data} and 
Fig.~\ref{fig:AMISA}).
The north-east part of the ring structure is clear but is confused with the central source.

\begin{figure}
\centerline{\includegraphics[width=7.8cm,height=7.8cm,keepaspectratio,angle=-90]{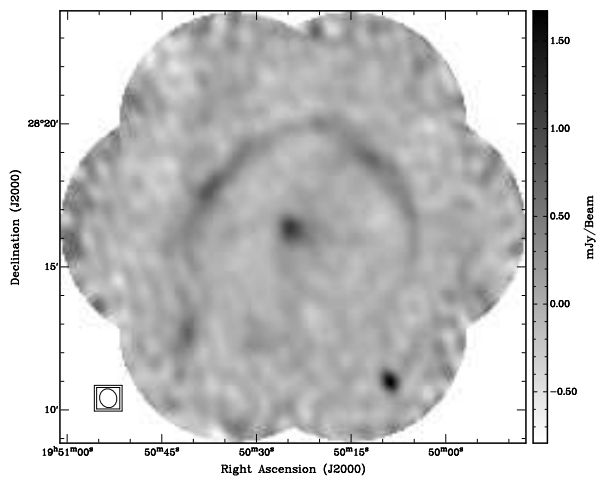}}
\caption{Map of a primary-beam-corrected combined-channel (14--18\,GHz) seven-point raster of \SNR(64.5)+(0.9)
observed by the AMI LA. The central
pointing has a thermal noise of $32~\mu$Jy~beam$^{-1}$ and the surrounding six pointings have a 
thermal noise of $58~\mu$Jy~beam$^{-1}$.\label{fig:LA-raster}}

\end{figure}

The AMI LA combined-channel raster map is shown in Fig.~\ref{fig:LA-raster}.
The ring is slightly resolved out by the AMI LA, but the source apparently near the centre
shows up clearly and has similar structure to that in the VLA maps at lower frequency. \textsc{fitflux}
was used to find the flux density of this near-central emission for each channel and these results are
listed in Table~\ref{tab:centreflux}. It is possible that the AMI LA
is slightly resolving out this object at the higher end of its frequency
coverage. However the slight drop in flux density is within the error bars.

\textsc{fitflux} was used to find the flux density of the whole object from the AMI SA combined-channel map,
and the flux density of the central source from the AMI LA combined-channel map.
These were $35.3\pm1.4$\,mJy and $2.49\pm0.10$\,mJy, respectively.
The latter was subtracted from the
former to produce an estimate of the flux density of the ring at 16\,GHz of $32.8\pm1.4$\,mJy.

\subsection{Data from the literature}\label{section:lit_data}

Most Galactic radio surveys do not possess the required resolution to resolve \SNR(64.5)+(0.9) well,
as it is only
8\arcmin\, in diameter. The Effelsberg 2.7-GHz survey data \citep{1984A+AS...58..197R} show
a small knot of emission
at the SNR's location, and a large amount of emission on broader angular scales. It was not possible
to extract meaningful flux densities for the ring and central source from these data.

This object is also covered by the Westerbork Synthesis Radio Telescope  Galactic plane
survey\footnote{http://www.ras.ucalgary.ca/wsrt\_survey.html}
(see e.g. \citealt{1996ApJS..107..239T}) at 327 MHz, which provides
low-frequency, large-scale information, at a resolution of 1\arcmin $\times 2$\arcmin$.11$\,at this declination.
Unfortunately the residual grating ring of a bright, distant source lies directly over the objects of interest.

Infrared data from the Infrared Astronomical Satellite (IRAS;\citealt{NASA RP-1190}) are too low in resolution to provide any
useful information about the region. The Second Digitised Sky Survey (DSS2)\footnote{http://archive.eso.org/dss/dss} shows only foreground stars in the Galaxy. The ATNF Pulsar Catalogue\footnote{http://www.atnf.csiro.au/research/pulsar/psrcat/} (see \citealt{2005AJ...129..1993}) shows no pulsars within one degree of the remnant.

\begin{table}
\caption{VLA and AMI LA I+Q flux densities of the emission close to the ring centre.\label{tab:centreflux}}
\begin{tabular}{cc} \hline
 $\nu$/GHz & $S_{\rm{i}}$/mJy \\ \hline
 1.43 & $17.07\pm0.36$ \\
 4.86   & $7.92\pm0.16$ \\
 8.46   & $4.01\pm0.08$ \\ 
14.2 & $2.69\pm0.17$ \\ 
15.0 & $2.58\pm0.12$ \\ 
15.7 & $2.50\pm0.13$ \\ 
16.4  & $2.48\pm0.12$ \\
17.1 & $2.27\pm0.12$ \\ 
17.9 & $2.02\pm0.11$ \\  \hline
\end{tabular}
\end{table}

\section{Discussion}\label{section:discussion}

For both the ring and the near-central emission we fit power-law spectra.
We define $\alpha$ such that flux density $S$
varies with frequency $\nu$ as $S\propto\nu^{-\alpha}$. 
Spectra were fitted to these data using a Gaussian likelihood function sampled by a
Markov Chain Monte Carlo technique. This method copes with asymmetric errors in $\log{S}$
across the frequency channels and provides an error estimate on
the spectral index directly from the posterior distribution.

Using the VLA 1.43\,GHz and AMI measurements of the flux density of the ring,
we calculate that it has a 
spectrum with $\alpha=0.47\pm0.03$. The structure and spectral index identify
the shell source \SNR(64.5)+(0.9) as a shell SNR. The surface brightness of the SNR at
1\,GHz is $\simeq3\times10^{-22}$W~m$^{-2}$\,Hz$^{-1}$\,sr$^{-1}$, which is faint for
known SNRs (in the faintest 10 per cent of identified SNRs - see e.g. \citealt{2004BASI...32..335G}).
This relatively low surface brightness suggests that the remnant is old, but on the other hand,
its highly circular shell suggests it is young. \citep{2006A&A...455.1053T} suggest a possible
distance of 11\,kpc, based on possibly related \textsc{hi} features, at which it
would have a diameter of $\simeq$26\,pc.
This is not the first SNR visible
in NVSS data which has been overlooked (see e.g. \SNR(353.9)-(2.0) identified by \citealt{2001MNRAS.326..283G}).

\begin{figure}
\centerline{\includegraphics[width=6.5cm,height=6.5cm,keepaspectratio,angle=-90]{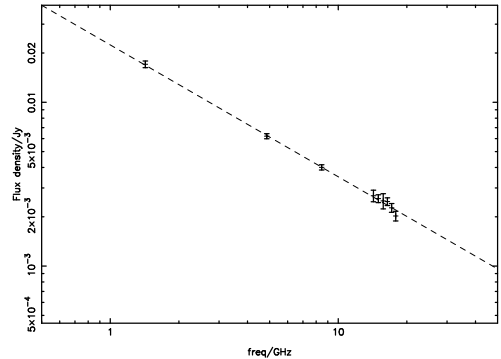}}
\caption{Logarithmic plot of the flux density measurements of the source seen near the centre of \SNR(64.5)+(0.9) in \ref{tab:centreflux}. The dashed line indicates a fitted
spectrum of $\alpha=0.81$ using the method described in Section~\ref{section:discussion}.\label{fig:centre_flux}}

\end{figure}

Fitting a spectrum to the data in Table~\ref{tab:centreflux},
we find that the central emission has a steep
spectrum with $\alpha=0.81\pm0.02$ (Fig. \ref{fig:centre_flux}). This emission is of interest,
because of both its apparent position relative to the ring and its structure.
It could be a background, unrelated, extragalactic source, or a Galactic source that may
or may not be related to the shell remnant.

First we consider a source at redshift $z$ around 0.1.
Then its angular size of around 1\arcmin\, implies a physical size D of 100\,kpc and its luminosity
$P_{\rm{1.43}}$ is around $3\times10^{22}$W\,Hz$^{-1}$sr$^{-1}$
(taking H$_{0}$ = 72\,km\,s$^{-1}$Mpc$^{-1}$).
These values are typical of FR-I \citep{1974MNRAS.167P..31F} 
twin-jet or tail radio sources powered by jet-producing machines. However the emission
is not reminiscent of known twin-jets or tails -- particularly at 8\,GHz -- and
the lack of spectral steepening from 1.43 to 18\,GHz\ also tells against it being
this type of radio source. A substantially higher redshift for the emission appears
ruled out: its structure is very difficult to reconcile with that of a powerful
radio source of whatever physical size or orientation, and the emission shows no sign of the
synchrotron or inverse-Compton losses associated with the extended regions of powerful sources.

`Halo' emission from certain galaxy clusters might resemble the shape of the near-central emission.
However, `halo' emission has an extremely steep radio spectrum (see e.g. \citealt{2008Natur.455..944B}).

The remaining extragalactic possibility would seem to be synchrotron emission from a nearby system
of interacting galaxies. At a redshift $z\simeq$0.02, its angular size of around 1\arcmin\,implies a physical
size $D$ of 20\,kpc and its flux density $S_{\rm{1.43\,GHz}}$ of 17.1\,mJy implies
a luminosity $P_{\rm{1.43\,GHz}}$ of $1\times10^{21}$WHz$^{-1}$sr$^{-1}$
Thus, this source's angular size, flux
density and structure are consistent with synchrotron emission from two or
three interacting galaxies at $z$ of around 0.01 -- 0.05.

However, from NVSS source counts\footnote{http://sundog.stsci.edu/first/catalog\_paper/node6.html},
we estimate at 15 per cent the probability of a source of $S_{\rm{1.43}}\geq$17\,mJy appearing to lie
inside the shell of the SNR purely by chance. Thus the probability of \emph{any} extragalactic radio source of this
flux density and of such large angular extent, lying within one arcminute of the centre, is minimal.

Turning to possible Galactic sources of emission for the near-central source, chance assocation
along the line-of-sight with
thermal Galactic sources such as \textsc{hii} regions is excluded both by the non-thermal radio
spectrum of the central source, and its lack of infra-red emission.
The location of the extended emission near the centre of the shell suggests
that \SNR(64.5)+(0.9) may be a `composite' SNR, i.e.\ a shell containing a
central pulsar wind nebula (PWN) -- for example, see \SNR(328.1)-(1.8)
\citep{2000ApJ...543..840D}. However, the radio spectral index of $\approx 0.8$
for the central source is considerably larger than expected, since pulsar wind
nebulae usually have relatively flat spectral indices at gigahertz frequencies
(e.g.\ \citealt{2008ApJ...687..516K}). If the central source is a pulsar wind
nebula associated with \SNR(64.5)+(0.9), then it requires a spectral break at
below 1.4~GHz, which would be unusual. A low frequency spectral break, at about
1.3~GHz -- three times lower than for any other PWN -- has been identified in
DA~495 ($=$\SNR(65.7)+(1.2), Kothes et al.). DA~495 is thought to be rather old
($\sim 20,000$~years), whereas the highly circular nature of the shell of
\SNR(64.5)+(0.9) suggests it is relatively young.

To identify the nature of this object and fully constrain its association
with the remnant, \textsc{hi} absorption-line observations are vital.

\section{Conclusions}

Follow-up obervations of an arc of emission seen in NVSS and CGPS data have been made
at frequencies from 1.4 to 18\,GHz. We draw the following conclusions from our observations:
\itemize{\item{\SNR(64.5)+(0.9) is confirmed as a shell SNR and its emission follows a power-law spectrum
with $\alpha=0.47\pm0.03$;}
\item{by considering its size, shape and spectral behaviour, and a probabilistic analysis using NVSS radio source counts,
we argue that the emission apparently at the centre of the ring is most unlikely to 
be any kind of extragalactic radio source;}
\item{it is difficult to avoid the conclusion that the apparently central emission is associated with the shell,
although its spectrum is very unlike that of a pulsar wind nebula, and it therefore merits further investigation, particularly
follow-up \textsc{hi} absorption line observations.}}

\section{ACKNOWLEDGMENTS}

We thank the staff of the Mullard Radio Astronomy Observatory for their invaluable
assistance in the commissioning and operation of AMI, which
is supported by Cambridge University and the STFC.
MLD, TMOF, CRG, NHW and TWS acknowledge the
support of PPARC/STFC studentships.

\bsp \label{lastpage}
\end{document}